\documentclass{jfm}
\usepackage{graphicx}
\usepackage{epstopdf, epsfig, amsmath}
\DeclareMathOperator{\sech}{sech}

\shorttitle{Coherent structure coloring}
\shortauthor{K.L. Schlueter-Kuck and J. O. Dabiri}

\title{Coherent structure coloring: identification of coherent structures from sparse data using graph theory}

\author{Kristy L. Schlueter-Kuck\aff{1}
  \corresp{\email{kristy@stanford.edu}}
  \and John O. Dabiri\aff{2}
  \corresp{\email{jodabiri@stanford.edu}}}

\affiliation{\aff{1}Department of Mechanical Engineering\\
Stanford University, Stanford, CA 94305, USA
\aff{2}Department of Mechanical Engineering \\
and Department of Civil and Environmental Engineering, \\
Stanford University, Stanford, CA 94305, USA}

\begin{document}

\maketitle

\begin{abstract}
We present a frame-invariant method for detecting coherent structures from Lagrangian flow trajectories that can be sparse in number, as is the case in many fluid mechanics applications of practical interest.  The method, based on principles used in graph coloring and spectral graph drawing algorithms, examines a measure of the kinematic dissimilarity of all pairs of fluid trajectories, either measured experimentally, e.g. using particle tracking velocimetry; or numerically, by advecting fluid particles in the Eulerian velocity field.  Coherence is assigned to groups of particles whose kinematics remain similar throughout the time interval for which trajectory data is available, regardless of their physical proximity to one another.  Through the use of several analytical and experimental validation cases, this algorithm is shown  to robustly detect coherent structures using significantly less flow data than is required by existing spectral graph theory methods.
\end{abstract}

\section{\label{sec:intro}Introduction}

The concept of coherence in fluid flows has historically been used to delineate packets of fluid elements that persist while the flow evolves without significant mixing with the surrounding fluid regions.  Coherence can frequently be visualized qualitatively by observing the evolution of passive tracers in a flow (e.g.~\citet{Haller2015},~\citet{Huhn2012}).  However, mathematical frameworks are needed to quantify such structures objectively.  Eulerian techniques for coherent structure identification include the q-criterion~\citep{Hunt1988}, $\lambda_2$ criterion~\citep{Jeong1995}, and the Okubo-Weiss parameter~\citep{Okubo1970, Weiss1991}.  All of these methods are frame-dependent, however.  Frame invariance is an important characteristic of a method for determining coherent structures; if a method identifies a structure boundary in one frame of reference, but not in another (for example, in a rotating reference frame), then the method may not be self-consistent in its characterization of fluid coherence~\citep{Haller2005}.  

As an alternative, Lagrangian techniques have also been developed, many based on analysis of the deformation gradient tensor of the flow field~\citep{Haller2000b, Shadden2005}.  These methods use information regarding the trajectories of fluid particles, as opposed to their velocity or acceleration, which ensures frame invariance since relative particle position does not depend on the reference frame in which it is measured.

Despite the vast array of current applications, identification of coherent structures based on the deformation gradient has some limitations as well.  For instance, knowledge of a discretized version of the entire flow field is typically required, of the sort obtained from computational analysis or particle image velocimetry (PIV).  However, some common empirical tools for fluid flow measurement do not provide velocity data in the whole flow field.  One such technique, particle tracking velocimetry (PTV), (e.g.~\citet{Chang1984},~\citet{Racca1988}), is particularly useful in applications where velocity data in three dimensions is required, or where the entire flow cannot be densely and uniformly seeded, as in studies of ocean currents.  Particle tracking algorithms often result in much sparser velocity measurements than techniques such as PIV.  For example,~\citet{Davis1991} reviews a wide range of studies utilizing artificial ocean drifters to study nominally two-dimensional ocean surface flows, where the number of drifters ranges from 14 to 300 per study.  In three dimensions, a number of PTV studies have utilized between 800 and 5000 particle trajectories~\citep{Virant1997, Luthi2005, Murai2007, Kim2013}.  Recently, several PIV and PTV techniques have been developed to increase the density of data collected~\citep{Elsinga2006,Schanz2016}.  Despite these advances, in many situations seeding the flow with a large number of particles can be extremely difficult, e.g. in situ measurements of atmospheric and oceanic flows that rely on naturally occurring particulate fields~\citep{Hong2014, Katija2008}.  Hence, the need for techniques to analyze sparse data persists.

Deformation-gradient based methods for identification of coherent structures fail for sparse trajectory data due to the assumptions inherent in analyzing the deformation gradient tensor, $\frac{\partial\boldsymbol{x}}{\partial\boldsymbol{X}}$, where $\boldsymbol{x}=\boldsymbol{x}(\boldsymbol{X})$ maps the initial location of a fluid element, $\boldsymbol{X}$, to its location $\boldsymbol{x}$ at a later time.  The principal assumption that is no longer satisfied is the initial close separation of flow trajectories, since the trajectory spacing cannot be controlled a priori.  Moreover, the determination of the finite time Lyapunov exponent (FTLE) field requires linearization of the flow map~\citep{Haller2000b}, which also breaks down for well-spaced flow trajectories.  Therefore, an alternate approach is needed to extract coherent structures from sparse velocity data.

Several methods have been developed recently in an attempt to address this issue, a number of which are reviewed by~\citet{Allshouse2015}.  One such method is based on braid theory~\citep{Allshouse2012}, which maps two dimensional fluid particle trajectories into three dimensions, where time is the third dimension.  Plotted in this way, the entwinement of trajectories with each other can be analyzed, and surfaces surrounding sets of trajectories that do not grow with time can be identified, indicating the presence of coherent structures.  While the braid method is useful for two-dimensional datasets, its extension to higher spatial dimensions has not yet been achieved.  Additionally, the braid-based analysis can become quite complex and computationally expensive for large numbers of trajectories.

A second method developed for use with sparse data sets is the cluster-based approach by~\citet{Froyland2015}, and it is well suited for PTV datasets due to its ability to handle both sparse and incomplete fluid particle trajectories.  The method uses the Euclidean distance from each particle to the center of each of a predetermined number of clusters to assign to each fluid particle a probability of belonging to each cluster while simultaneously determining the location of the cluster centers.  This is accomplished using the iterative fuzzy c-means algorithm developed for use in cluster-theory~\citep{Bezdek1984}.  While the authors proved that this method can accurately detect coherent structures from a variety of benchmark flows, in addition to global ocean drifter data, it has the distinct disadvantage that the number of clusters be known a priori.  

In order to address the need to determine the number of clusters a priori in the cluster-based approach of~\citet{Froyland2015},~\citet{Hadjighasem2016} recently developed a method based on spectral graph theory.  This method relies on the concept of a graph, or a set of nodes connected by edges, where in this case, the nodes represent Lagrangian particles, and the edges are weighted by the inverse of the average distance between particle pairs.  By examining the smallest magnitude eigenvalues associated with the generalized eigenvalue problem of the graph Laplacian, the authors developed a heuristic for determining the number of coherent structures in the flow.  The authors use this information as input to the K-means clustering approach to determine the centers of the coherent structures in the flow.  The test-cases used by the authors in validating this approach demonstrate its effectiveness in identifying coherent structures without knowing the number of clusters a priori.  However, for most flows analyzed, the number of trajectories used, on the order of tens of thousands, far exceeds the number of trajectories available for most PTV and ocean drifter datasets.  It is also important to note that the method for determining the number of coherent structures is heuristic and therefore difficult to generalize.  Other graph theory-based methods have also been developed recently to address the issues associated with current cluster and braid based approaches~\citep{Banisch2016}.

We propose an alternate graph theory-based metric that weights the edges not by the distance between corresponding particles, but by a metric of kinematic dissimilarity, regardless of spatial proximity.  This method is frame invariant because it does not consider particle velocity, only the spatial location of each fluid particle relative to other particles in the flow.  In analogy to graph coloring algorithms that partition nodes with large connecting edge weights, the present method solves an eigenvalue problem to partition fluid particle trajectories according to their kinematic dissimilarity. This approach can be considered an application of spectral graph drawing, which uses eigenvectors of matrices associated with a graph to visualize certain characteristics of the graph~\citep{Koren2005}.  The present method results in a coherent structure coloring (CSC) field, where similar values of CSC indicate regions of the flow that are coherent, according to the present definition.  In this way, all coherent structures in the flow can be identified simultaneously, without prior knowledge of their number.  Methods for extracting individual coherent structures from the CSC field are discussed.  The method was tested using three validation cases, including two canonical analytic flows: an oscillating quadruple-gyre and a Bickley jet; and one experimental dataset: a two-dimensional cross-section of a high stroke-ratio vortex ring.

The following section details the mathematical derivation of the algorithm used for coherent structure identification.  Section 3 presents the results of the three test cases described above and characterizes the sensitivity of the method to certain parameters, including the number and initial location of the particles.  Section 4 compares the results of the CSC method to the results of other graph theory-based methods.  Section 5 summarizes the results of the study and provides avenues for future development.

\section{Methods}

Coherence, defined here as the kinematic similarity of Lagrangian fluid trajectories, regardless of their spatial proximity, can be identified in flows with arbitrary time-dependence using a graph theory-based approach.  The graph $G$ is defined as the superset $G=(V,E,W)$, where $V$ represents the set of nodes in the graph, $E$ is the set of edges connecting the nodes, and $W$ are the weights corresponding to the edge set.  Assuming that the trajectories of a set of $N$ Lagrangian fluid particles is known at $T$ time steps, a graph representing the flow can be constructed, wherein each node represents a fluid particle.  Unlike previous methods that weight the edges of such a graph based on the proximity of the fluid trajectories (e.g. using Euclidian distance), here we use a weight based on kinematic dissimilarity.  We hypothesize, and later demonstrate, that coherent structures can be identified more robustly by quantifying the extent to which fluid trajectory kinematics are different, rather than the extent to which fluid particle trajectories remain in proximity over time.  To this end, each edge, representing the connection between a pair of particles, is weighted by the standard deviation of the distance between the two fluid trajectories over their duration, normalized by the average distance between the fluid particle trajectories during the same period.  The edge weights can be represented numerically using the weighted adjacency matrix $A$, where $a_{ij}$ contains the weight of the edge connecting particle $i$ and particle $j$:
\begin{equation} \label{eq:diss_weight}
a_{ij}=\frac{1}{\overline{r_{ij}}T^{1/2}}\left[\sum_{k=0}^{T-1}(\overline{r_{ij}}-r_{ij}(t_k))^2\right]^{1/2}
\end{equation}
where $r_{ij}(t_k)$ is the distance between two particles $i$ and $j$ at time $t_k$, and $\overline{r_{ij}}$ is the average distance between the two fluid particle trajectories.

Graph coloring is a labeling of nodes in a graph such that node pairs with large edge weights are assigned dissimilar values~\citep{Munoz2005}.  This makes graph coloring a natural tool for coherent structure identification based on the kinematic dissimilarity metric in equation~\ref{eq:diss_weight}.  We pose this as the one-dimensional problem of maximizing
\begin{equation}
z=\frac{1}{2}\sum_{i=1}^N\sum_{j=1}^N\left(x_i-x_j\right)^2a_{ij}
\end{equation}
where $N$ is the number of rows and columns in the weighted adjacency matrix $A$ (i.e. the number of particles) and $X$ is a row vector containing the value of coherent structure coloring (CSC) associated with each particle.    By maximizing $z$ we are in effect determining CSC values such that fluid particle trajectories that are kinematically dissimilar (i.e. where the weight of the edge between them $a_{ij}$ is large) are assigned CSC values that are as different as possible.  Following~\citet{Hall1970}, we define the degree matrix $D$, which contains the row sums of the adjacency matrix along the diagonal, as
\begin{equation}
d_{ij}= \left\{
\begin{array}{ll}
    0,			& i\neq j\\
    \sum_{k=1}^N a_{ik},       & i=j.
\end{array} \right.
\end{equation}
We also define the graph Laplacian, $L=D-A$.  The maximization problem can then be manipulated as follows:
\begin{eqnarray}
z & =&\frac{1}{2}\sum_{i=1}^N\sum_{j=1}^N\left(x_i-x_j\right)^2a_{ij}\\
  & =&\frac{1}{2}\left(\sum_{i=1}^Nx_i^2\sum_{k=1}^Na_{ik}-2\sum_{i=1}^N\sum_{j=1}^Nx_ix_ja_{ij}+\sum_{j=1}^Nx_j^2\sum_{m=1}^Na_{mj}\right)\\
    & =&\sum_{i=1}^Nx_i^2\sum_{k=1}^Na_{ik}-\sum_{j=1}^N\sum_{i=1}^Nx_ix_ja_{ij}\\
    & =& X^\prime LX
\end{eqnarray}

In order to avoid the trivial case where $x_1=x_2=\dots=x_N=0$, and to bound $X$ to finite values, the constraint $X^\prime DX=1$ is imposed (another finite constraint can be imposed without loss of generality).  Given this constraint,  the maximization problem can be written in Lagrangian form as $z=X^\prime LX-\lambda(X^\prime DX-1)$.  As a necessary condition for $z$ to be a local maximum, $\frac{dz}{dX}=0$, yielding $2LX-2\lambda DX=0$, which can be written as
\begin{equation} \label{eq:gen_eig}
LX=\lambda DX
\end{equation}
which is the generalized eigenvalue problem for the graph Laplacian.  The generalized eigenvector $X$ that maximizes $z$ is the eigenvector corresponding to the maximum eigenvalue of equation~\ref{eq:gen_eig}~\citep{Hall1970}.  Each element of $X$ assigns that value of CSC to the corresponding fluid particle at the final time of the interval over which particle trajectories were compared.  The CSC vector can be mapped to the space of the original flow with arbitrary dimensionality based on the spatial locations of the particles.  Interpolation between the particles facilitates construction of the corresponding CSC field.  Thus, regions in the flow with a similar value of coherent structure coloring indicate coherence.

\section{Results}
The effectiveness of the coherent structure coloring algorithm is demonstrated using three example flows.  The first, a quadruple gyre, is an extension of the double gyre, which is frequently used in vortex detection algorithm verification~\citep{Allshouse2015, Froyland2015}.  Both the steady and the unsteady cases are examined.  The second verification case is the Bickley jet, which introduces complexities due to the presence of multiple coherent vortices as well as a meandering jet.  Finally, we apply the CSC method to sparse trajectories derived from a particle image velocimetry (PIV) dataset of a long stroke ratio vortex ring, where secondary and tertiary rings in addition to a trailing jet form behind the primary vortex ring.  This shows the robustness of the algorithm to errors associated with experimental data.  For the two analytical validation cases, a fifth-order, Runge-Kutta method was used to determine fluid trajectories.  For the PIV dataset, particle velocities were determined using linear interpolations between velocity vectors and particles were advected using the Euler method.

\subsection{Quadruple Gyre}
First, we examine the characteristics of the coherent structure coloring algorithm using the analytical quadruple gyre flow.  This flow is defined by
\begin{eqnarray}
\frac{dx}{dt} & =&-\pi A\sin(\pi f)\cos(\pi y)\\
\frac{dy}{dt} & =&-\pi A\cos(\pi f)\cos(\pi y)(2ax+b)
\end{eqnarray}
where $x$ and $y$ are the spatial coordinates, $t$ is time, and
\begin{eqnarray}
a =\epsilon\sin(\omega t), 
b =1-2\epsilon\sin(\omega t), 
f =ax^2+bx.
\end{eqnarray}

\begin{figure}
\includegraphics[width=\linewidth]{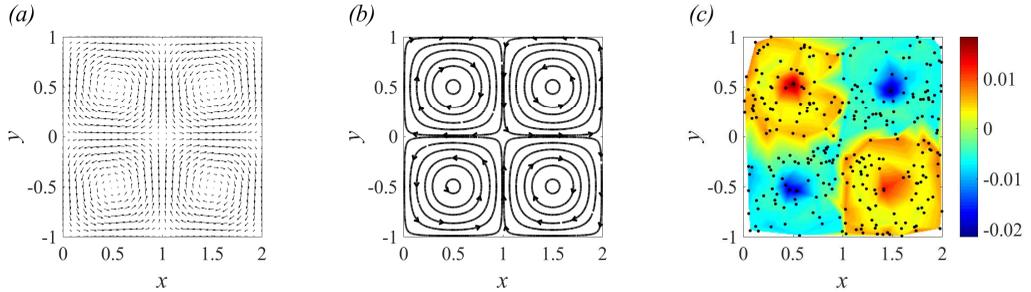}
\caption{Steady quadruple gyre flow (a) velocity vector field (b) streamlines (c) coherent structure coloring using 300 particles, particle locations indicated by black dots}
\label{fig:steady_quad}
\end{figure}

Here we examine two parameter cases: the steady case where $A=0.1$ and $\epsilon=0$; and the unsteady case where $A=0.1$, $\epsilon=0.1$, and $\omega=2\pi/10$.  Figure~\ref{fig:steady_quad} shows the velocity vector field, streamlines, and coherent structure coloring for the steady quadruple gyre, tracking only 300 particles over 40 time units.  
CSC is able to clearly delineate the four quadrants of the flow, and assigns a high value to the gyre centers.  Gyres in opposite corners have approximately the same value of coloring, due to their identical rotational orientation (clockwise in the upper left and lower right, and counterclockwise in the upper right and lower left quadrants).  This is a result of having a measure of coherence that does not conflate kinematic similarity and physical proximity.  The latter does not necessarily imply the former, and vice versa.  Large weights in the present adjacency matrix correspond to fluid particles that are kinematically dissimilar, and given that the weights correspond to the standard deviation of the distance between two particles~\textit{divided} by their average distance, the mean distance between particles and the standard deviation in their distances both contribute to coherence as defined by this algorithm.

\begin{figure}
\includegraphics[width=\linewidth]{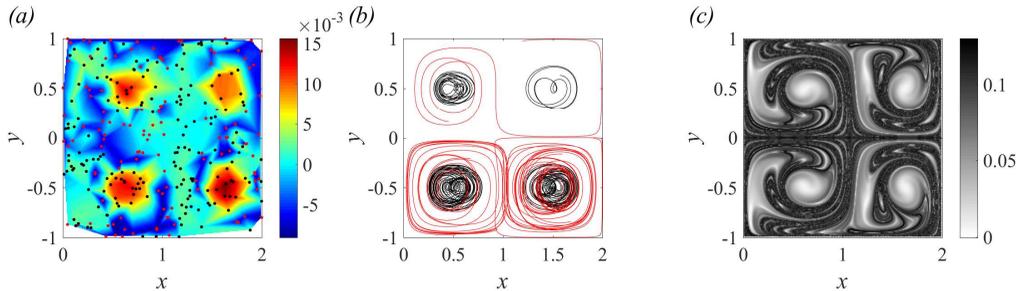}
\caption{Unsteady quadruple gyre, $\epsilon=0.1$, $A=0.1$, $T=[2.5,42.5]$ (a) coherent structure coloring using 300 particles, black dots show final locations of particles that remained in their initial quadrant, red dots show final locations of particles that switched quadrants during the coherent structure calculation time interval, (b) Fluid trajectories for particles with highest (black) and lowest (red) CSC values, (c) FTLE field, calculated over the time interval $T=[2.5,42.5]$, using 65,000 particles }
\label{fig:unsteady_quad}
\end{figure}

\begin{figure}
\includegraphics[width=\linewidth]{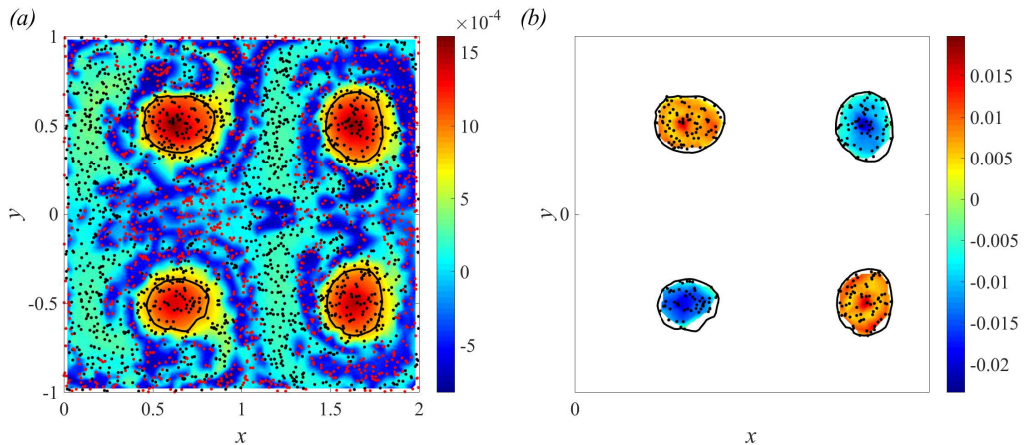}
\caption{Unsteady quadruple gyre, $\epsilon=0.1$, $A=0.1$, $T=[2.5,42.5]$, 3000 total particles (a) coherent structure coloring, black dots show final locations of particles that remained in their initial quadrant, red dots show final locations of particles that switched quadrants during the coherent structure calculation time interval.  Black lines trace contours of CSC=0.0009, (b) coherent structure coloring, algorithm applied recursively only to particles whose initial CSC value was greater than 0.0009}
\label{fig:quadgyre_recursive}
\end{figure}

The flow becomes significantly more complex when periodic oscillatory unsteadiness is introduced.  When comparing the coherence coloring to the FTLE of the same flow computed using 65,000 particles, seen in figure~\ref{fig:unsteady_quad}(c), it is clear that the largest positive value of coherence coloring correctly identifies all four vortex cores.  The negative values of CSC correspond with the regions in which particles have switched quadrants due to the oscillatory nature of the flow, as indicated by the red dots in figure~\ref{fig:unsteady_quad}(a).
In this case, the largest kinematic dissimilarity in the flow is between those particles that remain near the center of the quadrant in which they started, and those particles that switch quadrants.  This is highlighted by the particle traces shown in figure~\ref{fig:unsteady_quad}(b), which shows the trajectories of the particles with the largest positive value of coloring (in black) and those with the largest negative value of coloring (in red).  This result is in contrast to the steady case, in which the sign of vortex rotation was the predominant distinguishing feature.  Notably, the CSC algorithm can be applied recursively to the subset of particles with similarly high values of CSC in figure~\ref{fig:unsteady_quad}(a), in order to recover the vortex orientation information in figure~\ref{fig:steady_quad}(c).  This is demonstrated in figure~\ref{fig:quadgyre_recursive}.  
Here we see that for an initial CSC field calculated with 3000 particles, all four gyre centers are identified with a high CSC value.  In order to detect more subtle differences between the four gyre cores, a threshold value of $CSC>0.0009$ is set, indicated by the solid black lines surrounding the gyre cores in~\ref{fig:quadgyre_recursive}(a).  When the CSC algorithm is applied using only the particles exceeding this threshold value, the information distinguishing the gyres that rotate clockwise from those that rotate counterclockwise (as in the steady quadruple gyre case) is recovered.

For this dataset, a cluster-based method using fuzzy c-means clustering would correctly identify the four quadrants of the steady quadruple gyre as four separate coherent structures, if it is assumed a priori that there are four clusters present.  For the unsteady case, assuming the presence of four structures, the four gyre cores would again be correctly identified using this method.  However, if more than four coherent structures are assumed, the clustering method will detect the four cores and a number of additional structures, some of which may correspond to fluid parcels that did not switch quadrants but are not adjacent to the gyre cores~\citep{Allshouse2015}.  A braid based analysis for this flow would likely identify eight structures, again assuming an extension of the results of the double gyre system in ~\citet{Allshouse2015}.  In summary, the cluster based method requires a priori knowledge of the number of structures present, and the braid based analysis cannot be easily extended to three dimensions and is computationally expensive.  The CSC method addresses all of these issues.

\subsection{Bickley Jet}\label{sec:bickleyjet}
The Bickley jet, another analytical example, is frequently used as a model of zonal jets in the Earth's atmosphere~\citep{Rypina2007}.  It is a periodic flow comprising a spatially undulatory jet with counter rotating vortices above and below.  The flow is described by the stream function $\psi=\psi_0+\psi_1$, where
\begin{eqnarray}
\label{eq:bickley_eqn1a}
\psi_0 & =&c_3y-UL\tanh\left(y/L\right)\\
\label{eq:bickley_eqn1b}
\psi_1 & =&UL\sech^2\left(y/L\right)\sum_{n=1}^3\epsilon_n\cos\left(k_n\left(x-\sigma_nt\right)\right)
\end{eqnarray}

\begin{figure}
\centering
\includegraphics[width=\linewidth]{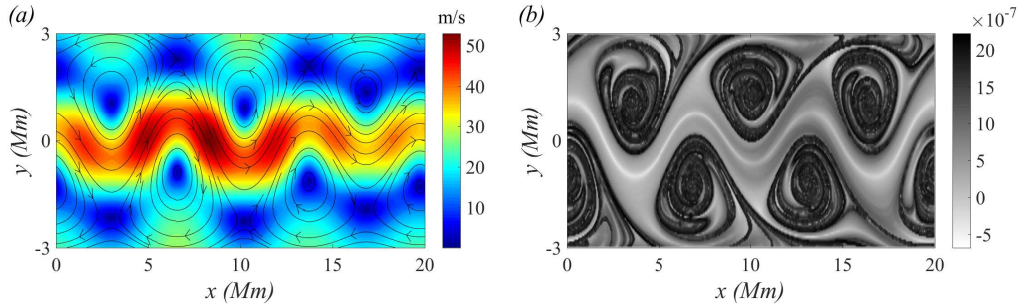}
\caption{Bickley jet, (a)  velocity magnitude with sample streamlines, (b) FTLE field, calculated over the time interval $t=[0$, $3456\times10^3]$ s using 48,000 particles} 
\label{fig:bickley1}
\end{figure}

We use similar values of the parameters as in~\citet{Hadjighasem2016}: $U=62.66$ ms$^{-1}$, $L=1770$ km, $k_n=2n/r_0$, $c=[0.1446U$, $0.205U$, $0.461U]$,  $\sigma=c-c(3)$, and $\epsilon=[0.0075$, $0.15$, $0.3]$, and the flow is computed on the interval $x=[0$, $20\times10^6]$ m, $y=[-3\times10^6$, $3\times10^6]$ m, over the time interval $t=[0$, $40]$ days, divided into 607 discrete time steps.  The flow was considered periodic in $x$.  For calculation of the CSC, particles were initialized randomly in the domain and advected with the flow.  The particles were followed over the entire time interval, even if they left the domain, analogous to how ocean drifters are tracked.  The velocity magnitude of the flow overlaid with streamlines is shown in~\ref{fig:bickley1}, along with the FTLE field calculated using 48,000 particles.

\begin{figure}
\centering
\includegraphics[width=\linewidth]{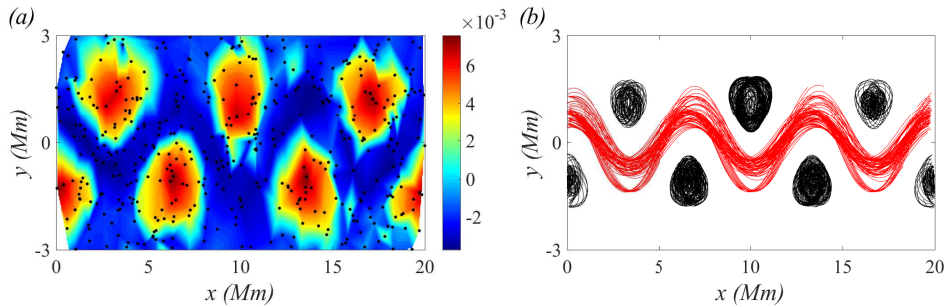}
\caption{Bickley jet, (a)  CSC contours overlaid with dots indicating final particle positions, 480 particles, $t=[0$, $3456\times10^3]$ s, (b) Particle tracks for particles with highest (black) and lowest (red) CSC values}
\label{fig:bickley2}
\end{figure}

Figure~\ref{fig:bickley2} shows the results of the coherent structure coloring algorithm using only 480 particles: (a) shows the CSC field overlaid with black dots indicating the final particle positions, and (b) shows particle tracks for the highest positive coloring values (in black) and the highest negative coloring values (in red).  
Without specifying the number of coherent structures a priori, the algorithm is able to accurately detect the centers of the three vortices above the jet and two full and two half vortices below.  However, if the seeding density was too low, such that one of the vortices contained no particles, or only a few, the structure could not be identified.  The jet itself is aligned with the most negative coloring contours, indicating that the largest kinematic dissimilarity detected is between the jet and the vortices flanking it.  It is noteworthy that the eigenvector associated with the largest eigenvalue of the generalized eigenvalue problem (i.e. the CSC) provides information about all of the coherent structures simultaneously, as opposed to $N$-cut approaches that recover one structure per eigenvalue among those selected heuristically.  This is in part because the present algorithm avoids unnecessarily restricting the definition of coherence to particles that are physically close; even the relative kinematics of particles that are far apart provides useful information regarding the coherent structures in the flow.

\begin{figure}
\centering
\includegraphics[width=\linewidth]{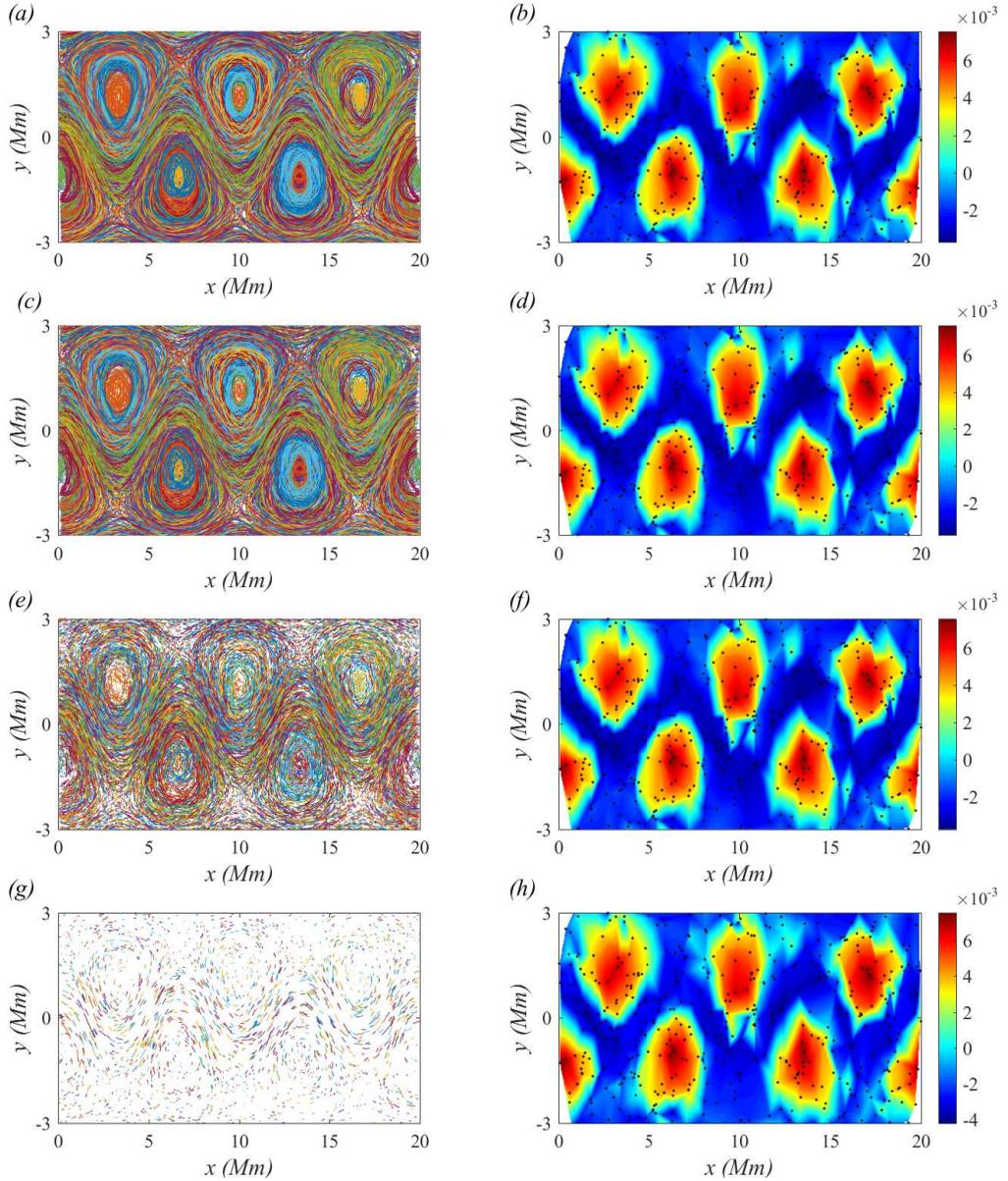}
\caption{Bickley jet, 480 particles (a) unbroken particle trajectories (b) CSC field for unbroken particle trajectories (c) particle trajectories, 10\% of position data deleted (d) CSC field for case where 10\% of particle position data is deleted (e) particle trajectories, 50\% of position data deleted (f) CSC field for case where 50\% of particle position data is deleted (g) particle trajectories, 90\% of position data deleted (h) CSC field for case where 90\% of particle position data is deleted.  Black dots indicate final particle position}
\label{fig:bickley_broken_480}
\end{figure}

The Bickley jet flow can also be used to characterize the robustness of the CSC method to broken or incomplete fluid particle trajectory information.  Two types of incomplete datasets are examined, each corresponding to a specific experimental circumstance in which data might be lost.  First, we examine the case in which a fluid particle trajectory is lost and later recovered, but is still identified as the same particle as before the data loss.  This could be the case for ocean drifters, e.g. when a drifter temporarily goes offline with associated data loss.  Additionally, certain PTV algorithms are capable of linking broken trajectories if data sets are sparse enough and particle trajectory crossings do not occur when breaks occur~\citep{Li2008}.  To characterize the response of the CSC method to this type of data loss, a dataset containing full trajectories of 480 particles was manipulated to randomly remove 10\%, 50\%, and 90\% of the trajectory data.  When considering the weight of a pair of particles, the CSC algorithm only considers the overlapping time interval in which both particles are present.  If a particle is not present during the time step for which the CSC field is computed, the trajectory information for that particle is not considered in the calculation of the adjacency matrix.  Hence, the size of the adjacency matrix is $n_p\times n_p$, where $n_p$ is the number of particles present in the domain during the time step in which the CSC field is calculated.  For this analysis, it was ensured that all 480 particles were present in the final time step so that their trajectory information could be used to calculate the CSC field.  This was done so that the analysis would show the effect of intermittent data loss as opposed to the effect of total number of particles.  The results are shown in figure~\ref{fig:bickley_broken_480}.
From this figure it is evident that in the case where pieces of particle trajectories can be linked and identified as broken pieces of the same trajectory, intermittent data loss does not adversely affect the robustness of the CSC algorithm, as long as there are a sufficient number of particles present in the time step for which the CSC field is calculated.  

\begin{figure}
\centering
\includegraphics[width=\linewidth]{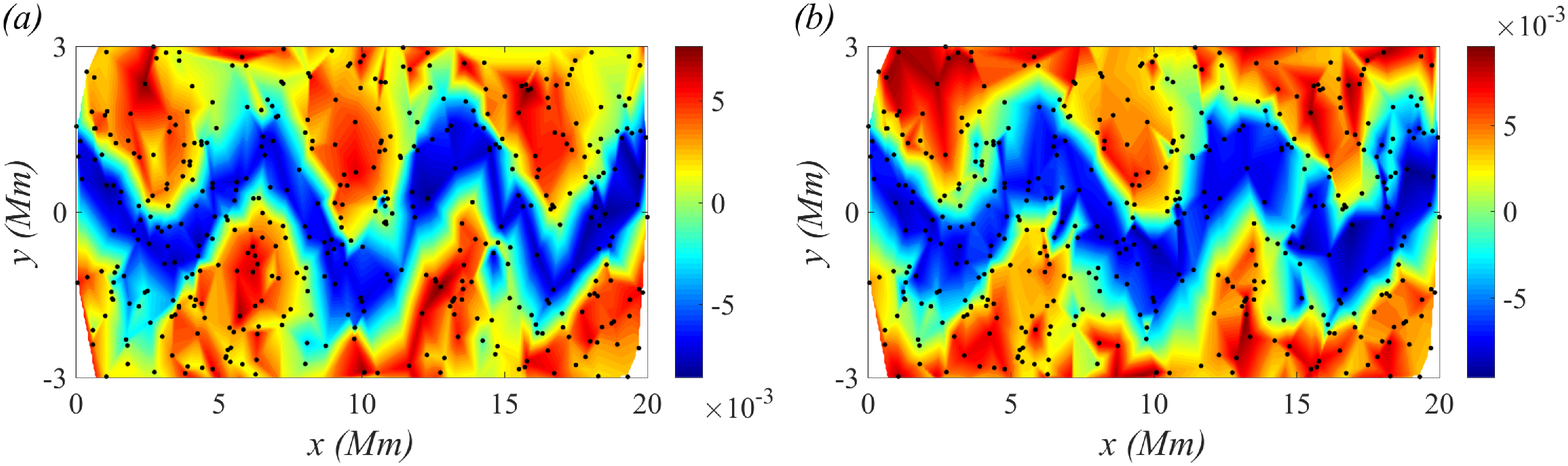}
\caption{Bickley jet CSC fields, 480 particles (a) average trajectory length of 2.6 eddy turnover times (b) average trajectory length of 1.7 eddy turnover times }
\label{fig:bickley_broken_480_v2}
\end{figure}

In order to characterize the CSC method in cases where broken trajectories cannot be reconstructed and the particle tracks must be treated as independent fluid particles, a baseline group of 480 particles was again examined.  A portion of the position data was then randomly deleted, and every time a break in the position data occurred, the remainder of the track was recharacterized as a separate particle trajectory.  The results from this analysis are shown in figure~\ref{fig:bickley_broken_480_v2}.
For the case with 480 unbroken particle trajectories, shown in figure~\ref{fig:bickley2}(a), all particles have a trajectory length that spans the entire time domain.  For the shorter particle trajectories shown in figure~\ref{fig:bickley_broken_480_v2}, it is evident that as the average particle trajectory length is shortened, the flanking vortices appear to blend together into two large coherent structures, one below the jet and one above.  This result can be understood by considering the length of the fluid particle trajectories relative to the eddy turnover time.  Based on the flow velocity around closed streamlines for the Bickley jet flow at $t=0$, the eddy turnover time is estimated to be approximately $279\times10^3$ s, and the time interval $t=[0$, $3456\times10^3]$ s is equivalent to approximately 12.4 eddy turnover times.  
Thus, while it is clear that the CSC algorithm is capable of handling broken trajectories of this type, an average trajectory length of approximately 2.6 eddy turnover times, as in figure~\ref{fig:bickley_broken_480_v2}(a), is necessary to distinguish individual coherent structures.  Otherwise, there is not enough information is available to effectively characterize the flow, even if the total observation time is a larger multiple of the eddy turnover time.

\begin{figure}
\centering
\includegraphics[width=4in]{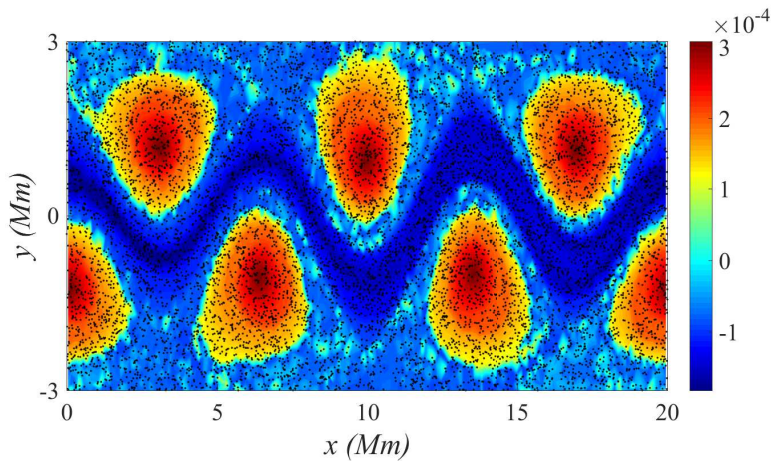}
\caption{Bickley jet CSC field, 12000 particles, black dots indicate final particle position}
\label{fig:bickley_highres}
\end{figure}

It is also useful to analyze and understand the response of the method to a large number of particles, approaching the quantity used for non-sparse methods such as FTLE analysis. As such, a CSC analysis of a Bickley jet seeded with 12000 particles was examined, and the resulting CSC field is shown in figure~\ref{fig:bickley_highres}.
The features of this CSC field are similar to what would be seen by clustering-based methods, if thresholding of the CSC were used to separate the vortex cores into distinct structures.  There are also similarities with what would be seen if vortices were extracted from the flow using the forward and reverse FTLE fields, including the five isolated vortex cores and two half cores.  Although not demonstrated here, the subsequent extraction of the coherent structures from the CSC field can be performed in a manner similar to the FTLE field analysis.  For example, one option is to use thresholding of the CSC field to determine boundaries of the coherent structures.  Additionally, the spatial gradients of the CSC field can be used to separate coherent structures from the background flow.  

In assessing the feasibility of high resolution CSC analysis, computational time is an important factor to consider.  Table~\ref{tab:run_times} provides a summary of computational run times on a single processor for the Bickley jet with three seeding densities.
{\setlength{\tabcolsep}{1em}
\begin{table}
  \begin{center}
\def~{\hphantom{0}}
  \begin{tabular}{lccc}\setlength\tabcolsep{24pt}
         & 480 particles  &2400 particles &12000 particles  \\[3pt]
         adjacency matrix calculation & 2.8 s    &79.1 s &   2345.2 s \\
	eigen-decomposition &   0.3 s  &  1.9s   & 213.2 s\\
  \end{tabular}
  \caption{Run times for Bickley jet flow on a single processor}
  \label{tab:run_times}
  \end{center}
\end{table}

\subsection{Vortex Ring}
Next, we examine a PIV dataset of a forming vortex ring with a high maximum stroke ratio.  The vortex ring is created in a water tank using a piston forced at speed $U$ a distance $L=Ut$ though a hollow cylinder of diameter $D$, which in turn ejects fluid from an axisymmetric nozzle with a sharp edge.  The shear layer formed inside the nozzle due to the motion of fluid through it rolls up at the nozzle exit forming a vortex ring.  If the maximum stroke ratio, $t^*_{max}=Ut_{max}/D$, where $t_{max}$ is the total time over which the piston is moving, is greater than 4, then a trailing jet and potentially secondary and tertiary vortex rings are formed behind the primary vortex ring~\citep{Gharib1998}.  The dataset examined here is of a vortex ring formed using a piston traveling with a constant velocity until a maximum stroke ratio of 8.  The Reynolds number based on the diameter is approximately 1800.  Details of the experimental setup and acquisition and processing of the PIV images for this dataset can be found in~\citet{Schlueter-Kuck2016}.

Figure~\ref{fig:vortexring1} (a) shows the vorticity field at $t^*=10.2$, after the piston has stopped moving, where $t^*=Ut/D$ is the nondimensional time, equivalent to the number of piston diameters that the piston has traveled.  
\begin{figure}
\centering
\includegraphics[width=\linewidth]{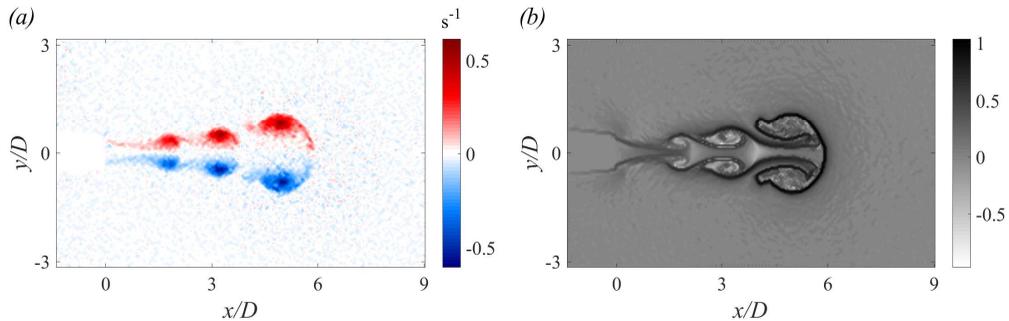}
\caption{Vortex ring, $t^*_{max}=8$, (a) vorticity field, $t^*=10.2$, (b) FTLE field, calculated over the time interval $t^*=[8.0$, $10.2]$, using 30,500 particles}
\label{fig:vortexring1}
\end{figure}
At this point, the leading vortex ring, as well as secondary and tertiary vortex rings and a trailing jet, are clearly visible.  The corresponding backward FTLE field, computed using 30,500 particles is shown in figure~\ref{fig:vortexring1} (b).  Figure~\ref{fig:vortexring2} shows the CSC calculated using a total of 150, 300, 600, and 2400 particles initiated at the nozzle exit plane near the left of the frame between $t^*=0.04$ and $t^*=8.4$.  The CSC algorithm identifies all three vortex rings, which is in qualitative agreement to the dark ridges of the FTLE field.
\begin{figure}
\centering
\includegraphics[width=\linewidth]{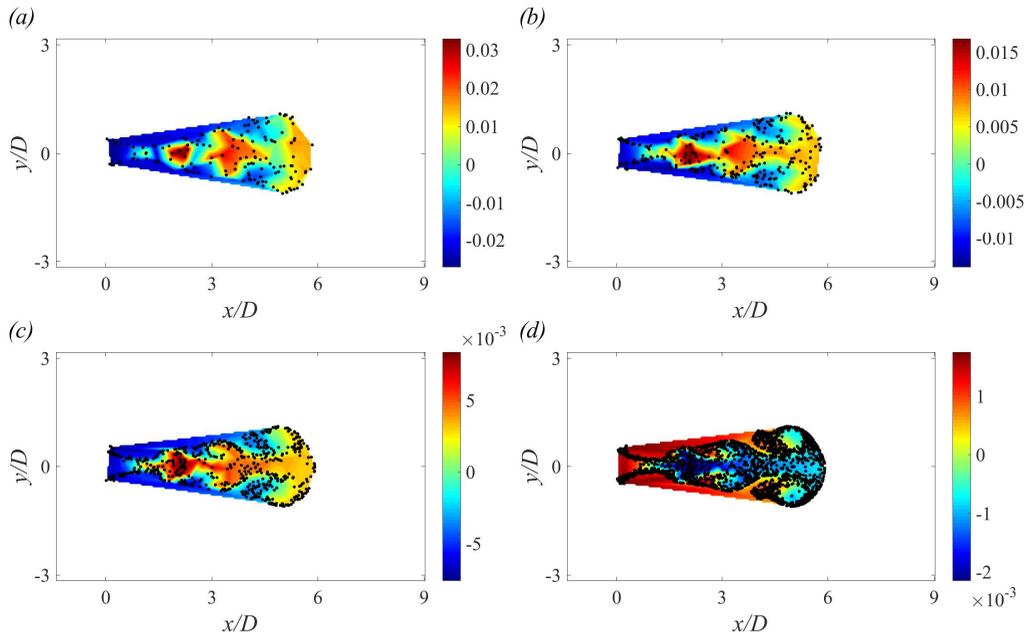}
\caption{Vortex ring CSC, for particles introduced at nozzle exit plane while vortex ring is forming, calculated over the time interval $t^*=[8.0$, $10.2]$, (a)  150 particles, (b) 300 particles (c) 600 particles, (d) 2400 particles}
\label{fig:vortexring2}
\end{figure}
While the resolution of the CSC contours increases with the number of particles, it is clear that 300 particles is sufficient to obtain a qualitatively similar result to the case with eight times as many particles, and to the FTLE calculation based on 30,500 particles.

The sensitivity of the CSC method to the size of the domain of particles and the time of release was also characterized using the vortex ring PIV data.
\begin{figure}
\centering
\includegraphics[width=\linewidth]{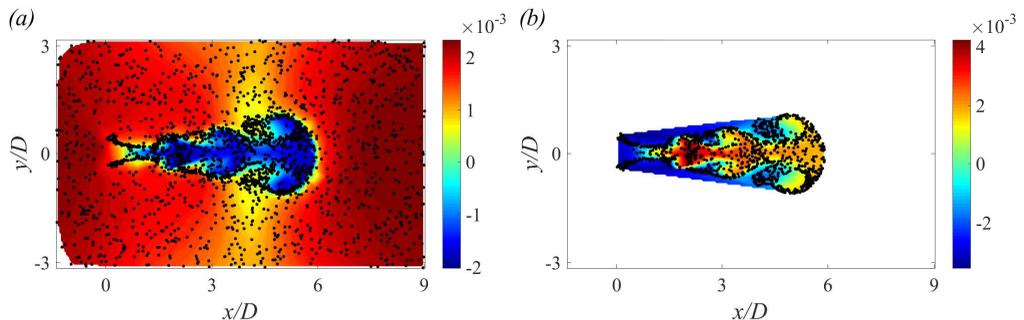}
\caption{Vortex ring CSC,  calculated over the time interval $t^*=[8.0$, $10.2]$ (a)  1200 particles initiated randomly in full domain at $t^*=0$ and 1200 particles introduced at nozzle exit plane during vortex ring formation time, $t^*=[0.04$, $8.4]$, (b) 1200 particles introduced at nozzle exit plane during vortex ring formation time, $t^*=[0.04$, $8.4]$}
\label{fig:vortexring3}
\end{figure}
In figure~\ref{fig:vortexring3}(a), the entire flow field was initialized with randomly located particles at $t^*=0$, and subsequently 1200 additional particles were added between $t^*=0.04$ and $t^*=8.4$ at the nozzle exit plane.  Because the CSC algorithm groups regions with a low normalized standard deviation in relative particle separation, the nominally quiescent background flow was assigned a CSC value that contrasts most sharply with the entire starting jet flow. Consequently, details of the internal structure of the vortex ring and trailing jet are lost.  However, if we recursively apply the CSC algorithm only to the flow trajectories in the starting jet, as shown in figure~\ref{fig:vortexring3}(b), we see that the algorithm is able to detect the structure of the primary, secondary, and tertiary vortex rings in greater detail, despite the fewer number of total particles.

\section{Comparison with other graph theory-based methods}
A related method for coherent structure detection that is also based on graph theory uses the concept of an eigen-gap heuristic to determine the number of coherent structures present in the flow~\citep{Hadjighasem2016}.  For this method, the weights assigned to the edges of the graph are equal to the reciprocal of the average distance between particle pairs.  The generalized eigenvalue problem solved in this method is $Lx=\lambda Dx$, where $L=D-A$ is the graph Laplacian, and $D$ is the diagonal degree matrix where $d_{ii}$ is equal to the sum of the elements in row $i$ of the adjacency matrix $A$.  This method assumes that by examining the smallest eigenvalues of the generalized eigenvalue problem, the number of coherent structures in a flow can be determined by locating the largest numerical gap between successive eigenvalues; the number of eigenvalues before the gap is assumed to correspond to the number of coherent regions in the flow.

Here we present an analysis of the aforementioned technique and its response to several input variables, comparing its robustness to the method introduced in this paper.  This analysis again uses the Bickley jet described by equations~\ref{eq:bickley_eqn1a} and~\ref{eq:bickley_eqn1b} and the values of the parameters listed in section~\ref{sec:bickleyjet}.  The response of the eigenvalue spectrum for the Bickley jet flow to changes in the initial position of tracer particles and to the value of the sparsification parameter $\epsilon$ are shown in figure~\ref{fig:eigengap1920} for 1920 particles and in figure~\ref{fig:eigengap480} for 480 particles.  For the case with 480 particles initialized randomly in the domain, the exact same particle trajectories were used as in the analysis in section~\ref{sec:bickleyjet} to allow for a direct comparison between the two methods.
\begin{figure}
\centering
\includegraphics[width=\linewidth]{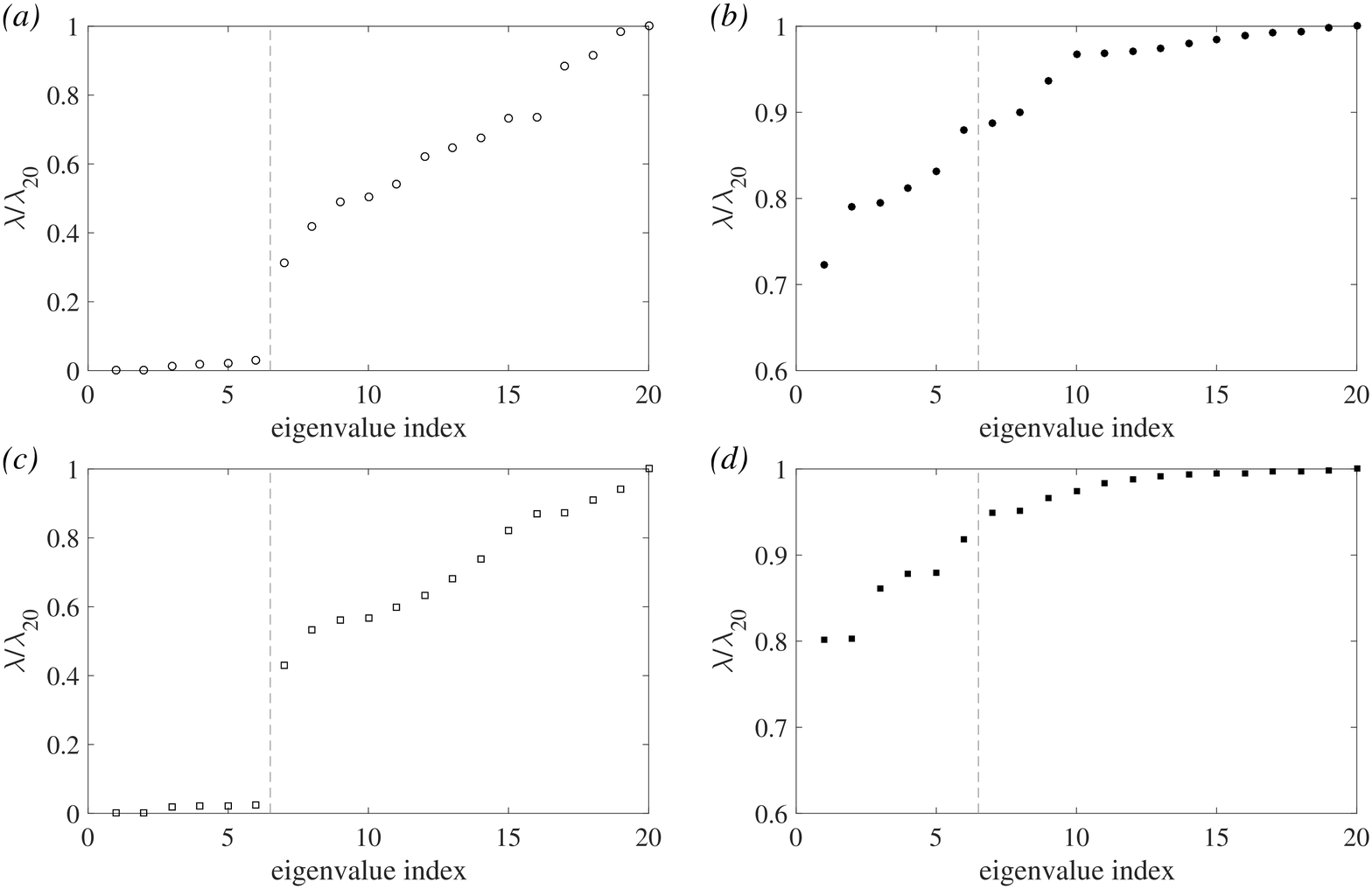}
\caption{Bickley jet eigenvalue spectra for 1920 particles (a) randomized particle initialization, sparsification of adjacency matrix for average particle pairwise distances greater than $3\times10^6$ m (b) randomized particle initialization, no sparsification of adjacency matrix (c)gridded particle initialization, sparsification of adjacency matrix for average particle pairwise distances greater than $3\times10^6$ m (d) gridded particle initialization, no sparsification of adjacency matrix.  Dotted vertical lines indicate the expected location of the eigen-gap based on six coherent structures}
\label{fig:eigengap1920}
\end{figure}
\begin{figure}
\centering
\includegraphics[width=\linewidth]{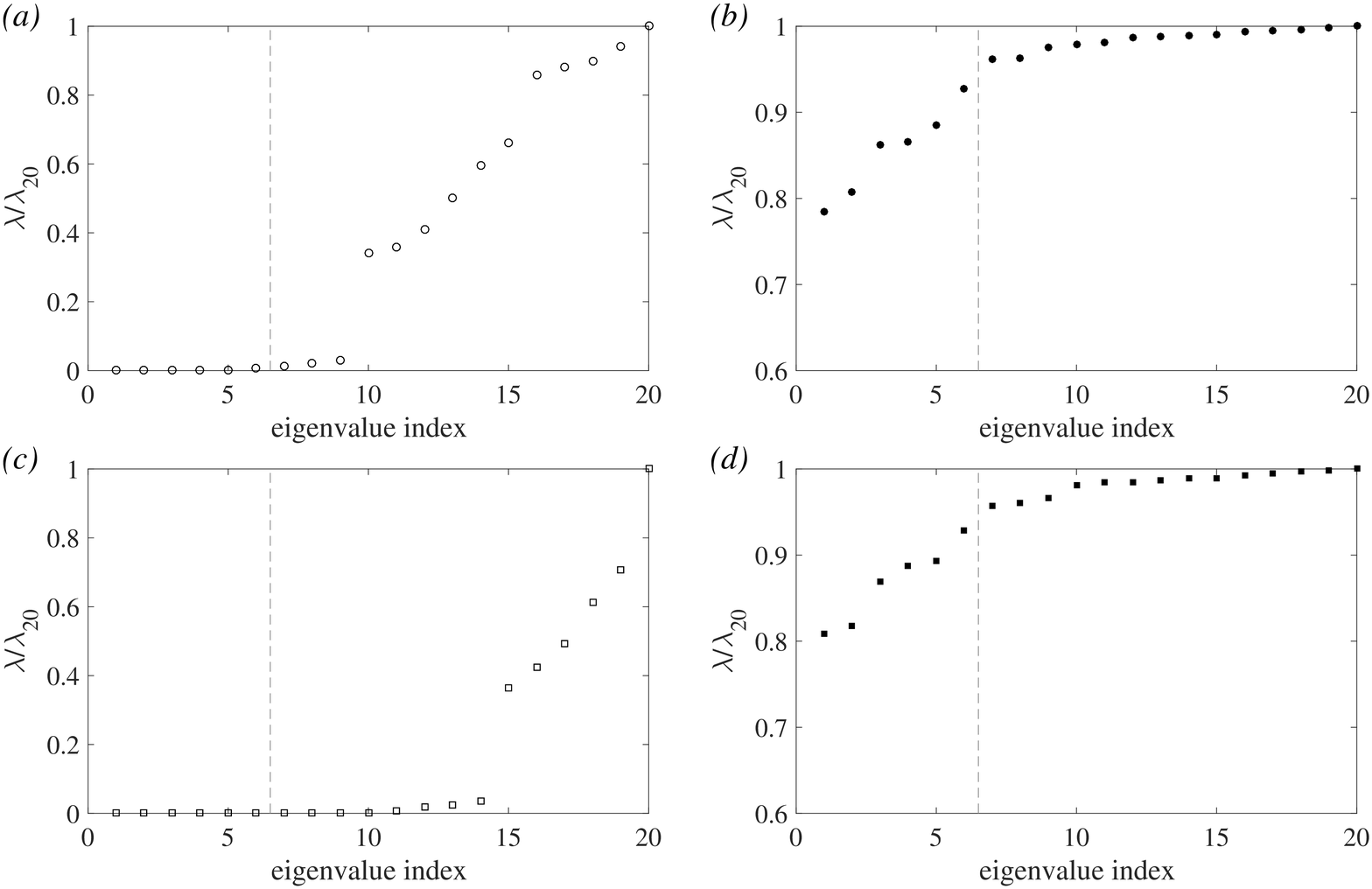}
\caption{Bickley jet eigenvalue spectra for 480 particles (a) randomized particle initialization, sparsification of adjacency matrix for average particle pairwise distances greater than $3\times10^6$ m (b) randomized particle initialization, no sparsification of adjacency matrix (c)gridded particle initialization, sparsification of adjacency matrix for average particle pairwise distances greater than $3\times10^6$ m (d) gridded particle initialization, no sparsification of adjacency matrix.  Dotted vertical lines indicate the expected location of the eigen-gap based on six coherent structures}
\label{fig:eigengap480}
\end{figure}

Based on prior knowledge of the Bickley jet flow, we expect to resolve six coherent structures: the five full vortices flanking the meandering jet, and due to the periodic nature of the flow in the x-direction, two half vortices identified together as a sixth coherent structure.  Thus, the eigen-gap heuristic should predict a numerical gap between the sixth and the seventh eigenvalues.  From figure~\ref{fig:eigengap1920}(a), (c) we can see that for 1920 particles, regardless of whether particles are initialized on a Cartesian grid or randomly throughout the domain,  the largest gap in the smallest 20 eigenvalues is between the sixth and seventh eigenvalues, as expected.  However, when the adjacency matrix is not sparsified to remove weights corresponding to particle pairs with an average distance greater than $3\times10^6$, as shown in figure~\ref{fig:eigengap1920}(b), (d), the eigen-gap is located between the first and second eigenvalues for the random particle initialization and between the second and third eigenvalues for the gridded particle initialization.  These results indicate that the eigen-gap heuristic is sensitive to the level of sparsification used in the adjacency matrix.  Consequently, without prior knowledge of the size of the coherent structures to inform an appropriate level of sparsification, this method is not able to robustly determine the number of coherent structures in the flow.

\begin{figure}
\centering
\includegraphics[width=\linewidth]{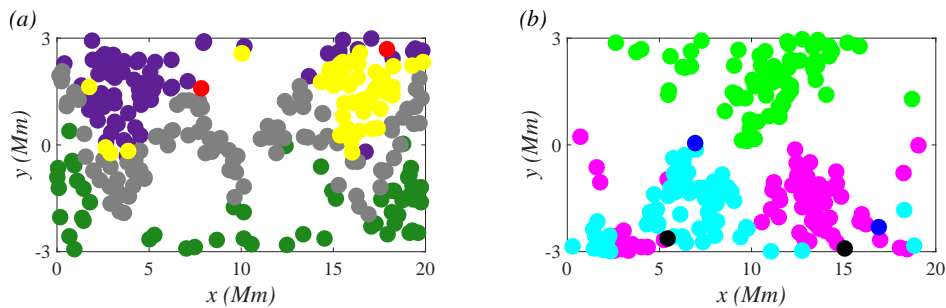}
\caption{K-means clustering of the Bickley jet with 480 randomly initialized particles, five of the ten total clusters identified are shown in (a) and the remaining five in (b)}
\label{fig:kmeans}
\end{figure}
\begin{figure}
\centering
\includegraphics[width=\linewidth]{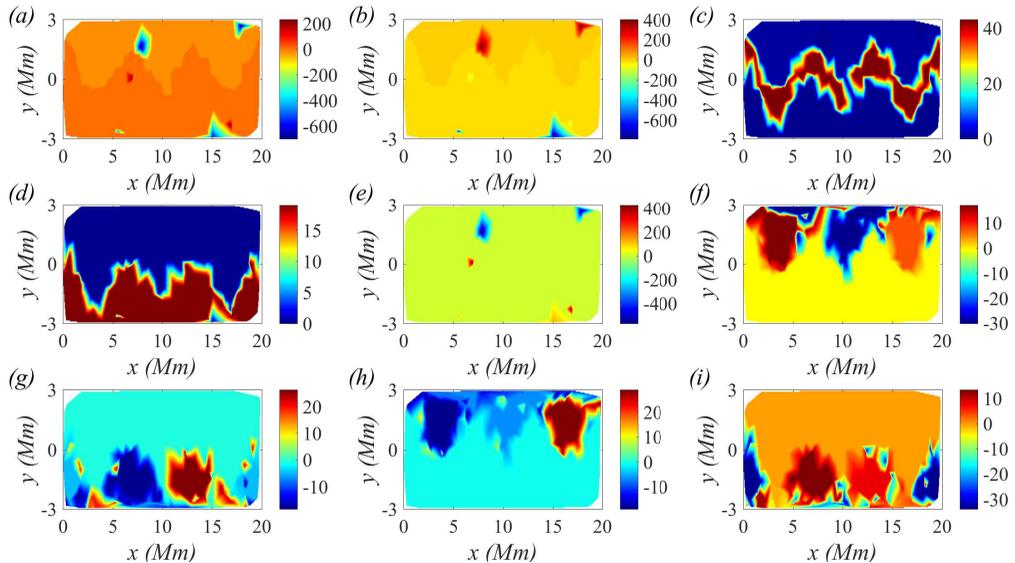}
\caption{Eigenvectors corresponding to the eigenvalues below the eigen-gap for the Bickley jet flow with 480 randomly initialized particles}
\label{fig:hallereigvs}
\end{figure}

The analysis of the eigenvalue spectrum for 480 particles, the same number used in the analysis of coherent structure coloring for the Bickley jet flow in section~\ref{sec:bickleyjet}, is shown in figure~\ref{fig:eigengap480}.  Here we see in figure~\ref{fig:eigengap480}(a) that for random particle initialization with sparsification, the eigen-gap is between the ninth and tenth eigenvalues.  Additionally, if the particles are initialized on a grid (figure~\ref{fig:eigengap480}(c)), or sparsification is not used (figure~\ref{fig:eigengap480}(b)), or both (figure~\ref{fig:eigengap480}(d)), the eigen-gap is also not correctly identified.  Based on this analysis, we can conclude that for small numbers of tracer particles, on the order of $10^2$ to $10^3$, use of the eigen-gap heuristic to determine the number of coherent structures in the flow is ineffective based on the lack of robustness to initial tracer locations (often beyond the control of the investigator for experimental applications) and to the level of sparsification of the adjacency matrix (which requires a priori knowledge of the size of the coherent structures, if sparsification is to be used effectively).

Despite an incorrect identification of the eigen-gap, the coherent structures can theoretically still be identified using a K-means clustering of the eigenvectors associated with the eigenvalues below the eigen-gap according to this method.  To be sure, if the eigen-gap heuristic overestimates the number of structures, some structures might be split into several, and if the number of structures is underestimated, several structures might be merged together.  Thus, for the case where 480 particles were randomly initialized in the domain and sparsification was used in the analysis (i.e. figure~\ref{fig:eigengap480}(a)), the results of the K-means clustering is shown in figure~\ref{fig:kmeans}, and the nine eigenvectors used in the clustering analysis are shown in figure~\ref{fig:hallereigvs}.  The clustering analysis searched for ten groups corresponding to the nine coherent structures expected from the eigen-gap heuristic, in addition to one structure for the incoherent background flow.  In figure~\ref{fig:kmeans}, these ten clusters are plotted between two separate panels to aid in visualization of the individual clusters.  From the clustering, we can see that the gray cluster roughly corresponds to the meandering jet, while the flanking vortices are somewhat consistent with the purple, yellow, and light green clusters on the top and the dark green, cyan, and magenta clusters on the bottom.  The black, red and blue clusters identify only a few seemingly random particles each.  Even if the three smallest clusters are ignored, the boundaries of the clusters corresponding to the vortices are significantly different from the boundaries of the vortices themselves, as observed in the FTLE analysis in figure~\ref{fig:bickley1}(b).  From observing the eigenvectors used for this analysis, it is evident that the first, second, and fifth eigenvectors (figure~\ref{fig:hallereigvs}(a),(b),(e)) are responsible for the small clusters identified by the K-means analysis, while the remaining six eigenvectors roughly correspond to the boundaries of groups of the flanking vortices and meandering jet.  Although not shown, if K-means clustering is performed using the third and fifth through ninth eigenvectors to identify seven clusters, the clusters identified are almost identical to the clusters in figure~\ref{fig:kmeans} excluding the small blue, red, and black clusters.  However, the boundaries of these clusters are still not consistent with the boundaries of the jet and the vortices.

\section{Conclusions}\label{sec:conclusions}
This paper presents an algorithm for detecting coherence in flows where only sparse velocity data is available, as is often the case in particle tracking velocimetry, or oceanographic tracking of surface floats.  In this regime, alternative methods evaluating coherence either require knowledge of the number of coherent structures a priori, or break down due to the sparsity of the data.  The present method, based on the concepts of graph coloring and spectral graph drawing, examines the kinematic dissimilarity of every pair of trajectories, and organizes this data into a weighted adjacency matrix.  As such, we consider a different definition of coherence from other coherent structure detection methods, which only consider groups of particles that remain close as time progresses without mixing with the surrounding fluid to be coherent structures. The eigenvector associated with the maximum eigenvalue of the generalized eigenvalue problem $LX=\lambda DX$ assigns a value of coherent structure coloring to each particle, such that similar CSC values indicate coherence in the flow.  This algorithm has several inherent strengths, including that the number of coherent structures does not need to be known a priori.  Additionally, information about all coherent structures in the flow is contained in a single eigenvector of the generalized eigenvalue problem associated with the graph Laplacian.  The algorithm is also capable of detecting coherent structures with the small number of trajectories associated with many PTV and ocean drifter datasets, and was shown to be robust to different types of data loss common in particle/drifter tracking applications.  Although only two dimensional datasets were examined here, the kinematic dissimilarity metric in equation~\ref{eq:diss_weight} and the subsequent maximization problem can both be extended to higher dimensions without loss of generality and with limited additional computational cost, since the adjacency matrix scales with the square of the number of particles, independent of the dimensionality of their trajectories.  The CSC method has the potential to be extended to analyze other properties of fluid flows in addition to coherence, and it may also find application in other data analysis problems for which coherent structure identification remains a challenge.

A MATLAB implementation of the CSC algorithm is available for free download at http://dabirilab.com/software.

\begin{acknowledgments}
This work was supported by the U.S National Science Foundation and by the Department of Defense (DoD) through the National Defense Science $\&$ Engineering Graduate Fellowship (NDSEG) Program.
\end{acknowledgments}

\bibliographystyle{jfm}
\bibliography{Schlueter}

\end{document}